# Skin supersolidity slipperizing ice

Xi Zhang, Chang Q Sun

Xi Zhang,[1,2,a] Yongli Huang,[3,a] Zengsheng Ma[3], Yichun Zhou[3,] and Chang Q Sun[1,2,3*]

[1] *NOVITAS, School of Electrical and Electronic Engineering, Nanyang Technological University, Singapore 639798*

[3] *Center for Coordination Bond and Electronic Engineering, College of Materials Science and Engineering, China Jiliang University, Hangzhou 310018, China*

[3] *Key Laboratory of Low-dimensional Materials and Application Technology (Ministry of Education) and Faculty of Materials, Optoelectronics and Physics, Xiangtan University, Xiangtan, 411105, China*

Ecqsun@ntu.edu.sg

[a] X.Z. and Y.H. contribute equally.

Abstract

Consistency between theory predictions and measurements and calculations revealed that the skin of ice, containing water molecules with fewer than four neighbours, forms a supersolid phase that is highly polarized, elastic, hydrophobic, with ultra-low density and high thermal stability. The supersolidity of skin sliperizes ice.

Ice surface is abnormal [1-3], which is most slipper of known [4, 5]. The slippery was commonly perceived as a result of friction-heating or pressure-depressed melting. However, neither of them can explain why ice can be so slippery even while one is standing still on it. Faraday [6] postulated in 1850's that a thin film of liquid water covers the surface even at temperature below freezing to serve as lubricant. Investigations suggested that ice surface pre-melting happens as the vibration amplitudes of the surface atoms were measured folds greater than the bulk [4]. However, an interfacial force microscopy and a spherical glass probe investigation revealed the opposite [4]. The surface layer is viscoelastic at temperatures over the ranging from -10 to -30 °C revealed that, resulting from the absence of the liquid layer at very low temperatures. Therefore, the concept of surface pre-melting seems in conflicting with the ice-like nature of ultrathin films of water. MD simulations [7] suggested that freezing preferentially starts in the subsurface of water instead of the top surface layer that remains disordered during freezing. Furthermore, the bulk melting is mediated by topological defects that preserve the coordination of the tetrahedral network. Such defects form a region with a longer lifetime [8].

Recent work [9-11] confirmed that the H-O contraction, core electron entrapment and non-bonding lone-pair polarization result in the high-elasticity, self-lubrication, and low-friction of ice surface and the hydrophobicity of water surface, of which the mechanism is the same to that of metal nitride [12, 13] and oxide [14] surfaces. The slippery or low-friction of ice surface [5] as results from the lone pair weak yet elastic interaction and the high density of surface charge instead of the liquid lubrication. Furthermore, because of the cohesive energy gain of the two intramolecular O-H bonds, a monolayer of water performs solid like with high elasticity and charge density because of the increase of molecular cohesive energy that raises the $T_m$. This expectation coincides with higher surface charge density measured using thin film interferometry [15]. The strong surface field induced by the surface charge establishes a more ordered hydrogen-bonding network that promotes the forming of thicker water lubrication film between hydrophilic solid surfaces.

Figure 1 compares the residual Raman spectra of the $\omega_H$ of water with that of ice collected using the Glancing angle Raman spectroscopy by Donaldson and co-workers [3]. The identical frequency of 3450 $cm^{-1}$ for the H-O stretching confirms the skin supersolidity of water and ice. The raw data were collected from water at room temperature and from ice (larger angle at -20 °C and smaller angle at -15 °C) at different angles between the surface normal and the reflected laser beam. Subtracting the spectrum collected from larger angle from the one collected at smaller angle upon spectral area normalization gives rise to the residual Raman spectrum. Molecular Dynamics (MD) calculation discriminates the skin from the bulk interior. In the skin region, water molecules become smaller but their separation is enlarged.

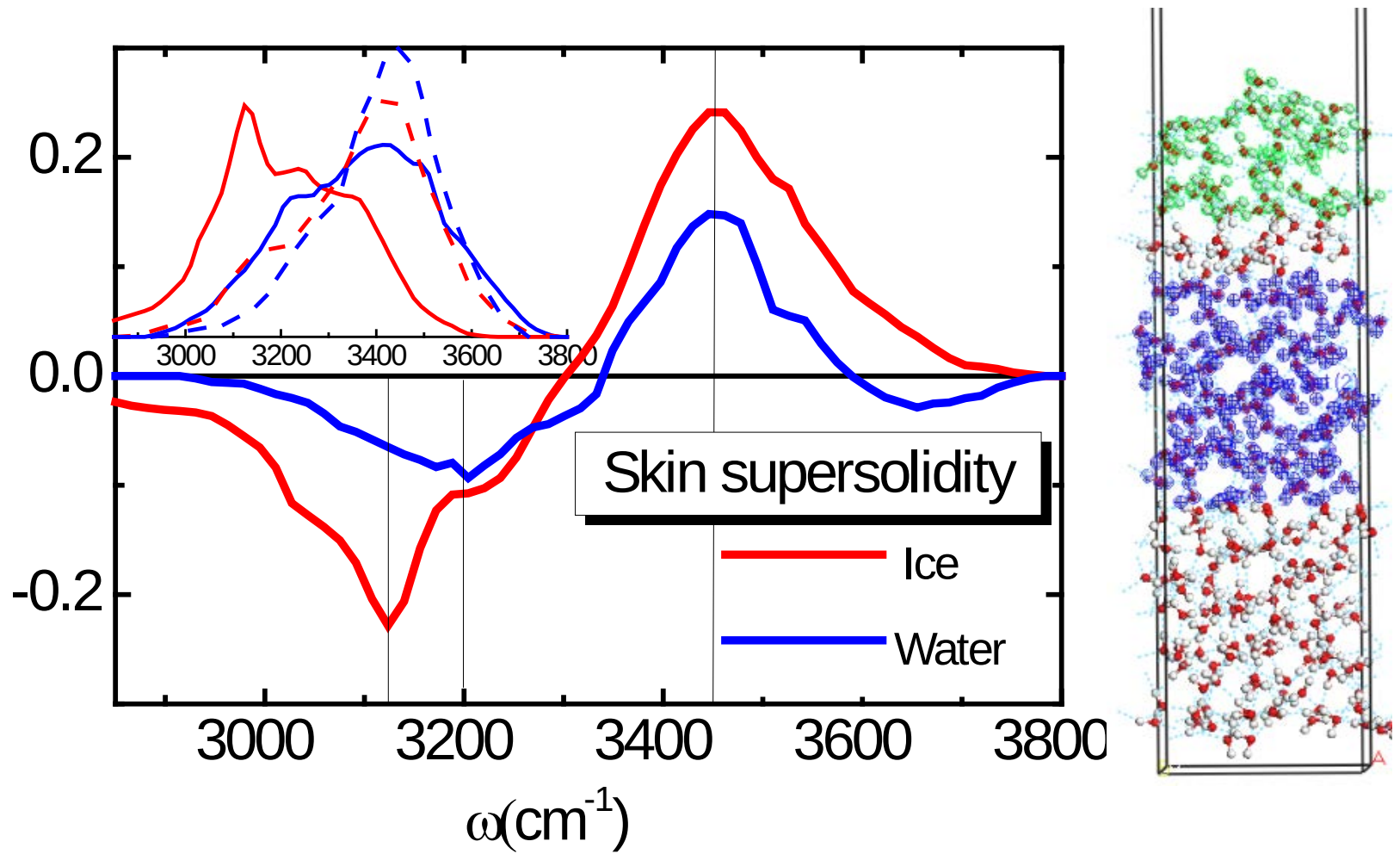

Figure 1 Residual Raman spectra of the O-H stretching modes of bulk water at room-temperature (solid blue trace), the air-water interface (dashed blue trace), bulk ice at -20 °C (dotted red trace), and the air-ice interface at -15 °C (dashed red trace) detected using glancing angle Raman spectroscopy with insets being the raw data of measurements. The residual Raman spectra were obtained by subtracting the spectrum collected at larger angles (between the surface normal and the reflection beam) from the one collected at small angles ([3]) upon the spectral area being normalized. MD calculation suggests that in the skin region, water molecules become smaller but their separation is enlarged.

Surface pre-melting is ruled out. As listed in Table 1, the skin of water and ice share the same $\omega_H$ = 3450 $cm^{-1}$ value for the H-O stretching vibration mode. The $\omega_H$ = 3200 $cm^{-1}$ for the bulk water and $\omega_H$ = 3125 $cm^{-1}$ for the bulk ice. In contrast, $\omega_H$ = 3650 $cm^{-1}$ for gases in vapor composed of dimers. Based on the derivatives in the recent work[16], density, $d_{OO}$, $d_x$, $E_x$, and density of each phase can be derived, see Table 1. According to the current notation, the $T_m$ is proportional to the bond energy of the H-O bond that becomes shorter and stronger at the surface skin.

Figure 2 shows the sampling procedure for extending the Ice Rule to the H-bond and the ideal structure of ice and water [11]. The central tetrahedron in Figure 1c illustrates the elegant Ice Rule of Pauling [17]. In the hexagonal or cubic ice phase the oxygen ions form each a tetrahedron with an O---O bond length 0.276 nm, while the O-H-bond length measures only 0.096 nm. Every oxygen ion is surrounded by four hydrogen ions and each hydrogen ion is connected to two oxygen ions. Maintaining the internal $H_2O$ molecule structure, the minimum energy position of a proton is not half-way between two adjacent oxygen ions. There are two equivalent positions that a hydrogen ion may occupy on the line of the O---O bond, a far and a near position. Thus a rule leads to the “frustration” of positions of the proton for a ground state configuration: for each oxygen ion, two of the neighboring protons must reside in the far position and two of them in the near, so-called “two-in two-out” frustration. The open tetrahedral structure of ice affords many equivalent states including spin glasses that satisfy the Ice Rule.

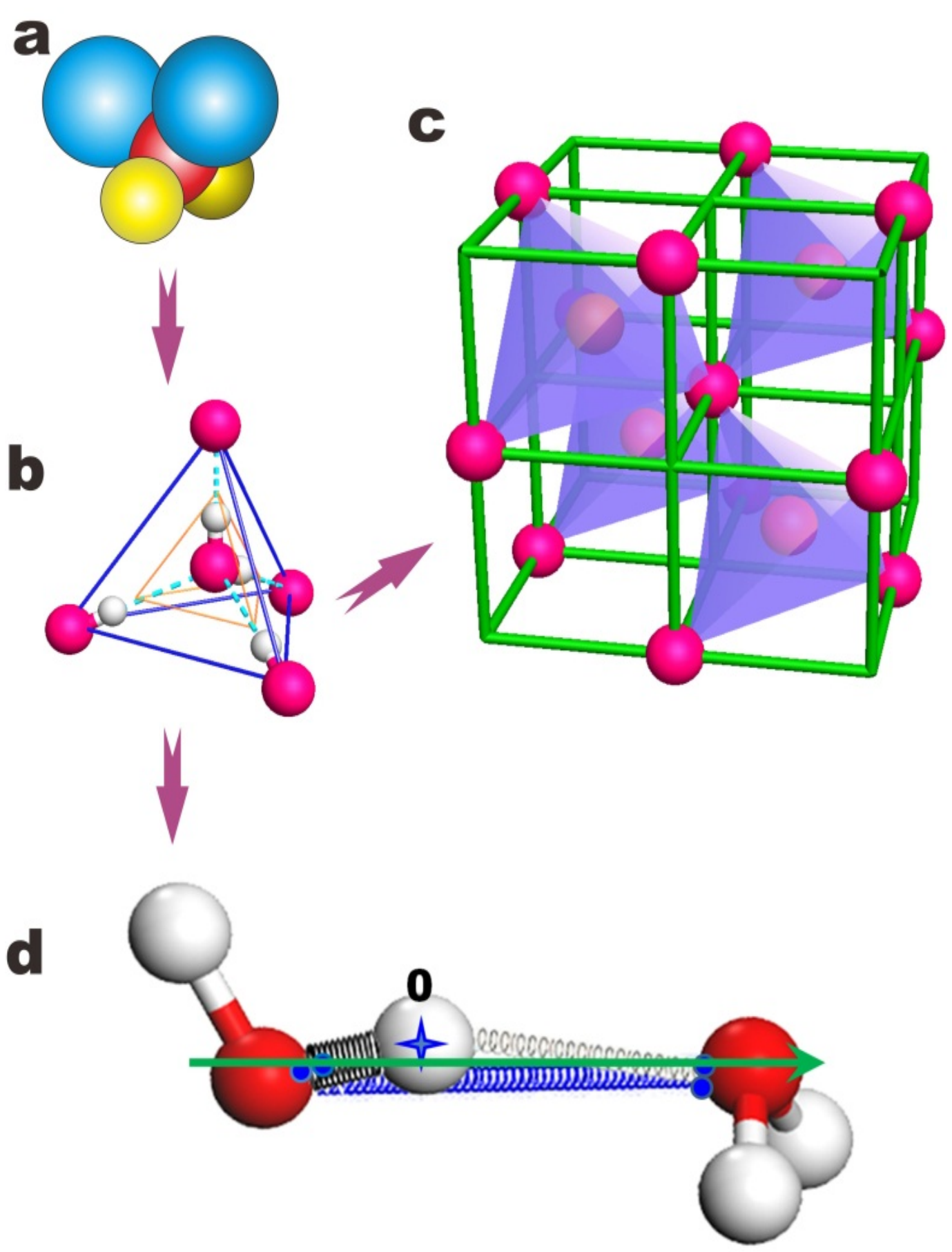

Figure 2 (a) Sampling procedure for extending Pauling’s Ice Rule [17]. The $sp^3$-hybrided oxygen with two lone pairs (green) and two bonding (yellow) orbits forms a quasi-tetrahedron of $C_{2v}$ group symmetry [18]. An extension of this quasi-tetrahedron yields an (b) an ideal tetrahedron of $C_{3v}$ that contains two $H_2O$ molecules and four identical O:H-O bonds. Packing the basic $C_{3v}$ blocks in an $sp^3$ order yields (c) a diamond structure that correlates the size, separation, and mass density of molecules packing in water and ice. (d) The H-bond forms a pair of asymmetric, coupled, H-bridged oscillators whose relaxation in length and energy mysterizes water and ice [11].

Undercoordination of water molecules shortens and stiffens the H-O bond and lengthens and softens the O:H is elongated, which lowers the density of packing. The stiffened H-O bond raise the frequency of vibration and deppens the potential well, resulting in O1s excessive energy shift to deeper [9].

Table 1 Shows the segmental bond length, vibration frequency, binding energy and density of water ice derived using the following relations[9, 10, 16, 19, 20]:

$$\begin{cases} d_{OO} = \sqrt{3}a/2 \;=\; 2.6950\rho^{-1/3} \\ d_L = \; 2.5621\times\left[1-0.0055\times exp\left(d_H/0.2428\right)\right] \end{cases}$$

$$\omega_x = \left(2\pi c\right)^{-1}\sqrt{\frac{k_x + k_C}{m_x}}$$

Where $k_c$ is the 2nd differential of the Coulomb potential and $k_x$ the 2nd differential of the respective short-rage potential for the H-O and O:H bond. The $m_x$ is the reduced mass of vibration dimers.

Table 1 Skin supersolidity ($\omega_x$, $d_{x,}$ $E_{x,}$ $\rho$) of water and ice derived from the measurements (indicated with refs) and calculated based on the H-bond model and structure model [16, 20].

| | Water (298 K) | | Ice ($\rho_{min}$) | Ice | Vapor |
|---|---|---|---|---|---|
| | bulk | skin | bulk | 80 K | dimer |
| $\omega_H$(cm$^{-1}$) | **3200**[3] | **3450**[3] | **3125**[3] | **3090**[10] | **3650**[21] |
| $\omega_L$(cm$^{-1}$)[10] | 220 | ~180[9] | 210 | 235 | 0 |
| $d_{OO}$(Å)[16] | **2.700**[22] | **2.965**[23] | 2.771 | 2.751 | **2.980**[23] |
| $d_H$(Å)[16] | 0.9981 | 0.8406 | 0.9676 | 0.9771 | 0.8030 |
| $d_L$(Å)[16] | 1.6969 | 2.1126 | 1.8034 | 1.7739 | ≥2.177 |
| ρ(g·cm$^{-3}$)[16] | 0.9945 | 0.7509 | **0.92**[24] | 0.94[24] | ≤0.7396 |
| $E_L$(meV)$\propto(\omega_x \times d_x)^2$ | 91.6 | 95[25] | 94.2 | 114.2 | 0 |
| *$E_L$(meV) ($q_H$ = 0.20 e) | 24.6 | 24.4 | 26.2 | 44.3 | 0 |
| $E_L$(meV) (0.17 e) | 33.4 | 33.8 | 35.1 | 52.0 | 0 |
| $E_L$(meV) (0.10 e) | 49.9 | Increase with the drop of $q_H$. | | | |
| $E_L$(meV) (0.05 e) | 58.3 | | | | |
| $E_H$(eV) $\propto(\omega_x \times d_x)^2$ | 4.4294 | 3.6518 | 3.97[10] | 3.9582 | 3.7300 |
| *$E_H$(eV) ($q_H$ = 0.20 e) | 3.6201 | 7.1967 | 4.0987 | 3.9416 | 8.6429 |
| $E_H$(eV) (0.17 e) | 3.6203 | 7.1968 | 4.0990 | 3.9418 | 8.6429 |
| $E_H$(eV) (0.10 e) | 3.6207 | Insensitive to $q_H$. | | | |
| $E_H$(eV) (0.05 e) | 3.6209 | | | | |

*Obtained by solving the Lagrangian motion equation[20].

In summary, water molecular undercoordination shortens and stiffens the H-O bond and meanwhile lengthens and softens the O:H bond. The shortening of the H-O bond raises the density of the core and the bond electrons, which in turn polarizes the nonbonding electrons. Therefore, the density of skin is lower (0.75 g·cm$^{-3}$). The high elasticity and the high density of dipoles form the essential conditions for the supersolidity [26], which slipperizes ice.

1. T. Ishiyama, H. Takahashi, and A. Morita, *Origin of Vibrational Spectroscopic Response at Ice Surface.* J Phys Chem Lett, 2012. **3**: 3001-3006.
2. X. Wei, P. Miranda, and Y. Shen, *Surface Vibrational Spectroscopic Study of Surface Melting of Ice.* Phys. Rev. Lett., 2001. **86**(8): 1554-1557.
3. T.F. Kahan, J.P. Reid, and D.J. Donaldson, *Spectroscopic probes of the quasi-liquid layer on ice.* J. Phys. Chem. A, 2007. **111**(43): 11006-11012.
4. R. Rosenberg, *Why ice is slippery?* Phys today, 2005(12): 50-55.
5. A.-M. Kietzig, S.G. Hatzikiriakos, and P. Englezos, *Physics of ice friction.* J. Appl. Phys., 2010. **107**(8): 081101-081115.
6. M. Faraday, *Experimental researches in chemical and physics*1859, London: Tayler and Francis

372.
7. L. Vrbka and P. Jungwirth, *Homogeneous freezing of water starts in the subsurface.* J. Phys. Chem. B, 2006. **110**(37): 18126-18129.
8. D. Donadio, P. Raiteri, and M. Parrinello, *Topological defects and bulk melting of hexagonal ice.* J. Phys. Chem. B, 2005. **109**(12): 5421-5424.
9. C.Q. Sun, X. Zhang, J. Zhou, Y. Huang, Y. Zhou, and W. Zheng, *Density, Elasticity, and Stability Anomalies of Water Molecules with Fewer than Four Neighbors.* J Phys Chem Lett, 2013. **4**: 2565-2570.
10. C.Q. Sun, X. Zhang, X. Fu, W. Zheng, J.-l. Kuo, Y. Zhou, Z. Shen, and J. Zhou, *Density and phonon-stiffness anomalies of water and ice in the full temperature range.* J Phys Chem Lett, 2013. **4**: 3238-3244.
11. C.Q. Sun, X. Zhang, and W.T. Zheng, *Hidden force opposing ice compression.* Chem Sci, 2012. **3**: 1455-1460.
12. C.Q. Sun, *Thermo-mechanical behavior of low-dimensional systems: The local bond average approach.* Prog. Mater Sci., 2009. **54**(2): 179-307.
13. C.Q. Sun, B.K. Tay, S.P. Lau, X.W. Sun, X.T. Zeng, S. Li, H.L. Bai, H. Liu, Z.H. Liu, and E.Y. Jiang, *Bond contraction and lone pair interaction at nitride surfaces.* J. Appl. Phys., 2001. **90**(5): 2615-2617.
14. C. Lu, Y.W. Mai, P.L. Tam, and Y.G. Shen, *Nanoindentation-induced elastic-plastic transition and size effect in alpha-Al2O3(0001).* Philos. Mag. Lett., 2007. **87**(6): 409-415.
15. S. Liu, J. Luo, G. Xie, and D. Guo, *Effect of surface charge on water film nanoconfined between hydrophilic solid surfaces.* J. Appl. Phys., 2009. **105**(12): 124301-124304.
16. Y. Huang, X. Zhang, Z. Ma, Y. Zhou, J. Zhou, W. Zheng, and C.Q. Sun, *Size, separation, structure order, and mass density of molecules packing in water and ice.* Sci Rep, Revised: http://arxiv.org/abs/1305.4246.
17. L. Pauling, *The structure and entropy of ice and of other crystals with some randomness of atomic arrangement.* J. Am. Chem. Soc., 1935. **57**: 2680-2684.
18. C.Q. Sun, *Oxidation electronics: bond-band-barrier correlation and its applications.* Prog. Mater Sci., 2003. **48**(6): 521-685.
19. Y. Huang, Z. Ma, X. Zhang, G. Zhou, Y. Zhou, and C.Q. Sun, *Asymmetric Potentials for the Length Symmetrization of Hydrogen Bond.* under review.
20. Y. Huang, X. Zhang, Z. Ma, Y. Zhou, G. Zhou, and C.Q. Sun, *Hydrogen-bond asymmetric local potentials in compressed ice.* J. Phys. Chem. B, Revised. **http://arxiv.org/abs/1305.2997**
21. Y.R. Shen and V. Ostroverkhov, *Sum-frequency vibrational spectroscopy on water interfaces: Polar orientation of water molecules at interfaces.* Chem. Rev., 2006. **106**(4): 1140-1154.
22. U. Bergmann, A. Di Cicco, P. Wernet, E. Principi, P. Glatzel, and A. Nilsson, *Nearest-neighbor oxygen distances in liquid water and ice observed by x-ray Raman based extended x-ray absorption fine structure.* J Chem Phys, 2007. **127**(17): 174504.
23. K.R. Wilson, R.D. Schaller, D.T. Co, R.J. Saykally, B.S. Rude, T. Catalano, and J.D. Bozek, *Surface relaxation in liquid water and methanol studied by x-ray absorption spectroscopy.* J. Chem. Phys., 2002. **117**(16): 7738-7744.
24. F. Mallamace, M. Broccio, C. Corsaro, A. Faraone, D. Majolino, V. Venuti, L. Liu, C.Y. Mou, and S.H. Chen, *Evidence of the existence of the low-density liquid phase in supercooled, confined water.* PNAS, 2007. **104**(2): 424-428.

25. M.W. Zhao, R.Q. Zhang, Y.Y. Xia, C. Song, and S.T. Lee, *Faceted silicon nanotubes: Structure, energetic, and passivation effects.* J Chem Phys C, 2007. **111**(3): 1234-1238.
26. C.Q. Sun, Y. Sun, Y.G. Ni, X. Zhang, J.S. Pan, X.H. Wang, J. Zhou, L.T. Li, W.T. Zheng, S.S. Yu, L.K. Pan, and Z. Sun, *Coulomb Repulsion at the Nanometer-Sized Contact: A Force Driving Superhydrophobicity, Superfluidity, Superlubricity, and Supersolidity.* J Chem Phys C, 2009. **113**(46): 20009-20019.